\documentclass[pre,aps]{revtex4}
\begin{document}
%\draft

\title{From Knowledge, Knowability and the Search for Objective Randomness to a New Vision of Complexity}
\author{Paolo Allegrini$^{1}$, Martina Giuntoli$^{2}$, Paolo Grigolini$^{2,3,4}$, Bruce J. West$^{5}$}
\address{$^{1}$ Istituto di Linguistica Computazionale del Consiglio Nazionale delle
Ricerche, Area della Ricerca di Pisa-S. Cataldo, Via Moruzzi 1, 56124, Ghezzano-Pisa, Italy }
\address{$^{2}$Center for Nonlinear Science, University of North Texas,
P.O. Box 311427, Denton, Texas, 76203-1427}
\address{$^{3}$Dipartimento di Fisica dell'Universit\'{a} di Pisa and INFM 
Piazza Torricelli 2, 56127 Pisa, Italy }
\address{$^{4}$Istituto dei Processi Chimico Fisici del CNR
Area della Ricerca di Pisa, Via G. Moruzzi 1,56124 Pisa, Italy }
\address{$^{5}$ Mathematics Division, Army Research Office, Research Triangle Park, NC 27709}
\date{\today}

\begin{abstract}
Herein we consider various concepts of entropy as measures of the
complexity of phenomena and in so doing encounter a fundamental
problem in physics that affects how we understand the nature of
reality. In essence the difficulty has to do with our understanding of
randomness, irreversibility and unpredictability using physical
theory, and these in turn undermine our certainty regarding what we
can and what we cannot know about complex phenomena in general. The
sources of complexity examined herein appear to be channels for the
amplification of naturally occurring randomness in the physical
world. Our analysis suggests that when the conditions for the
renormalization group apply, this spontaneous randomness, which is not
a reflection of our limited knowledge, but a genuine property of
nature, does not realize the conventional thermodynamic state, and a
new condition, intermediate between the dynamic and the thermodynamic
state, emerges. We argue that with this vision of complexity, life,
which with ordinary statistical mechanics seems to be foreign to
physics, becomes a natural consequence of dynamical
processes. 
\end{abstract}
\maketitle

\section{Introduction}

Why do things get more complicated with the passage of time?

While it may not be a mathematical theorem, it certainly seems clear
that as cultures, technologies, biological species and indeed most
large-scale systems, those with many interacting components, evolve
over time either they become more complex or they die out. The goal of
understanding the mechanisms by which evolution favors increased
complexity over time is too ambitious an undertaking for us here. We
do not explore these mechanisms in part because they are
phenomena-specific and we are concerned with only the universal
properties of complexity even though it has no clear definition. More
importantly, however, complexity is very often self-generating. Herein
we discuss many of the problems and paradoxes that are entangled in
the concept of complexity in the restricted domain of the physical
sciences. This is done because if complexity does have universal
properties they should be independent of the phenomena being studied
and therefore we choose the simplest context possible. For example the
generation and dissipation of fluctuations involved in complex
phenomena are examined through the concepts of irreversibility and
randomness, but more importantly through the \emph{ objectivity
principle}, see for example Monod \cite{monod}. The science of
complexity is multidisciplinary and so it might by argued that the
schema we construct, based on the paradigm of physics, is
incomplete. On the other hand, we believe that the principle of
objectiveness gives ``hard'' sciences like physics an important
advantage in addressing the difficult task of understanding complex
systems. As a matter of fact, we should not ignore centuries of
philosophy of science and, more generally speaking, epistemology, when
we refer to the concept of objectiveness. It is right in this case
indeed that science and philosophy benefit by mutual exchanges. We can
accept the fact that science is located in a better position in order
to understand and describe the reality of things, but the problem is
that we cannot still have a clear and complete definition of what we
mean by nature. Furthermore, we have to stress that in some cases
philosophers exert an important role of stimulus for scientists to
overcome the limitations of the current views. A significant example
is given by the paper of Pete Gunter, in these Proceedings
\cite{gunter}. This paper, in our opinion, sets challenges for the
scientists, and we shall discuss how to address some of these
challenges.  We have, in fact,
 three different problems
related to the principle of objectiveness. The
 first two are
connected to the scientist and the third one to nature itself. 
\begin{itemize}
\item
Since the scientists are human beings, while they investigate
the things around them, it happens that they not only describe the
facts the way they are, but they unavoidably corrupt and interact with the
things they are studying. This will become clearer below, when we shall 
explain the quantum mechanics paradigm. However, this is true also 
in a sense that is wider than the applicability of quantum mechanics itself.
\item
In addition to this, the scientist him/herself is part of a natural
system. For this reason it is possible that for what we call the {\em
vicious circle} the scientist cannot see all the things that he/she
should see. \footnote{In a certain way this reminds us of G\"odel and
his analysis of logical systems \cite{goedel}. If we consider his two
theorems on undecidability in closed formal systems in a wider way and
not strictly from a mathematical and philosophical point of view, then
it is easy to see that together with the impossibility of proving a
logical system "totally true from the inside", we have the uneasiness
of analysing a system that looks at itself from the inside expecting
to find the truth.} \footnote{This problem is also source of paradoxes
in quantum mechanics, but, even in the classical realm, it touches the
core problem of determinism in a strict sense, as well explained by
the ``egg's paradoxÓ , reported in
\cite{prigoginenouvellealliance}. This paradox, due to Diderot, refers
to the impossibility of classical Newtonian mechanics to describe the
development of living beings. In this paper we shall discuss some
recent developments that lead us to interpret complexity as a
condition intermediate between dynamics and thermodynamics, and make,
consequently, easier to re-conciliate life with statistical
mechanics.}
\item
Finally, we have to come back to the old and thorny problem concerning
the description of reality and the methods used in order to understand
it. Herein, we shall be mainly concerned with this latter problem,
namely the detection of true {\em unknowables} among many {\em
unknowns}, and how to deal with an unsatisfactory abundance of
unknowns, through a change of perspective, which is now becoming known
as the ``complexity paradigm''
\end{itemize}

 These three points are important to understand in order that we can
accept the principle of objectiveness even if some reserves have to be
made. Nevertheless the analysis and the criticism of the methods used
are a pre-eminent component of whichever process toward a renewed and
improved science. In spite of what we said, we , of course, value
science and scientists. Nobody would deny today that scientific
methods, reductionism included, make us able to describe the reality
of things to a very good approximation. The main purpose of what has
been said until now is just to "problematize" objectiveness, namely,
to question it and to consider whether it is either a cause or a
symptom of larger problems, especially within last century's
perspective, i. e. reductionism.  As a matter of fact, it is not at
all easy to define these two concepts, expecially reductionism. The
reason why this happens is because reductionism is a concept "in
fieri". We do not have a definition for reductionism that can satisfy
the whole scientific and philosophical community. One of the most
difficult aspects of the reductionism process to define is its
connection with the nature. In fact it is not clear if reductionism is
just a method used by the scientist and applied to nature in order to
understand it, or if it is an attribute of nature itself. Many
scientists and philosophers have written many articles and books about
this topics. Steven Weinberg, for example, has been very productive
from this point of view (see, e.g., his book "Dreams of a final
theory'' \cite{weinberg}). According to him, one would understand all
problems starting from fundamental equations governing particle
physics. Chemistry would be an application of physics, biology would
be a chemistry exercise, and so on.  We are not surprised that a
famous biologist, Ernst Mayr, was the first who clearly defined all
the different kinds of reductionisms \cite {mayr}, with the purpose of
circumscribing their applicability. According to his synthesis we can
classify reductionism in three categories. 
 
\begin{enumerate}

\item{Constitutive reductionism}, namely the scientific method in
which one studies the components to understand the system is no doubt
the key to success of decades of science, and no scientist doubts
this.  \item {Theory reduction}, or ``the phenomenon of relatively
autonomous theories becoming absorbed by, or reduced to, some other
more inclusive theory'' is, on the other hand, strongly criticized. A
classical project is the reduction of thermodynamics to mechanics,
which has never been accomplished.  We shall see later in this paper
that a new vision of complexity emerges from this failure in realizing
this project, in the case of non-ordinary statistical
mechanics.
\item {Explanatory reductionism}, at last, is completely
rejected by Mayr. This kind of reductionism states that a complete
knowledge of a system can be derived by the mere knowledge of its
components. In fact, it is not immediately and totally true that the
fact that we understand the single components make us able to
understand the whole system that we are considering.  \footnote{Even
the majority of physicists reject this kind of reductionism. John
Barrow \cite{barrow}, for example, in his ``Theories of
everything. The quest for ultimate explanation" makes the debate on
reductionism popular and within the reach of everybody. He actually
takes the example of a calculator and tries to understand if this
calculator can be considered just a certain amount of atoms or more
than that. This kind of process is very clear to all physicists who
deal with nonlinear science, the science where input and output are
not proportionally and constantly connected. This idea of
nonlinearity naturally leads to the theory of emergence.}
\end{enumerate}

We shall not try to
 resolve that controversy here, but we shall
indicate where we find
 that the new complexity approaches based on
nonlinear dynamics deviate
 from the traditional interpretation.
 
To appreciate how the complications of non-equilibrium phenomena have
changed our view of the world we go back a few decades to the work on
\emph {systems theory} pioneered by Von Bartalanffy \cite{von} and
subsequently developed by many scientists. Systems theory adopts the
perspective that determinism at best provides an inadequate
description of nature and a holistic approach is more suitable for
understanding phenomena in the social and life sciences. This
methodology casts the scientist in the role of ''problem solver'' so
that in order to extract information from the system the scientist
must develop a ''heuristic'' understanding of the problem to be solved
by means of metaphors. The holistic perspective assumes that the
scientific knowledge is universal in that laws within a given field of
study can often, if not always, be mirrored in all other fields of
study. We are often convinced that holism and reductionism are two
opposite concepts and that the first one belongs to philosophical
approach and the other to the scientific one. This is not totally
true, in fact there are many possibilities of compromising the two
points of view so they can peacefully coexist.

 The systems
theory approach to science has proved to be effective
 especially at
the interfaces of well-established disciplines, for
 example,
biophysics, biochemistry, information theory and cybernetics
 just to
name a few.  However, a metatheory involving the concept of
 complex
adaptive systems (CAS) has been invented and developed by
 members of
the Santa Fe Institute \cite{SFI}. According to Gell-Mann
\cite{gell} a CAS is a system that gathers information about itself
and its own behavior and from the perceived patterns which are
organized into a combination of descriptions and predictions,
modifies
 its behavior. Further, the interaction of such a CAS with
the
 environment provides feedback with which the survival
characteristics
 of the system are adjusted. This complicated
behavior leading to an
 internal change in the system associated with
decision making is not
 to be confused with the direct control
envisioned in cybernetics and
 other early forms of systems theory.

 To facilitate our discussion we adopt as a working definition of
complexity that found in Reulle's monograph, {\it Chance and Chaos}
\cite{ruelle}: ''Given a system and observer, the complexity of the
system is measured by the difficulty encountered by the observer in
extracting information from the system.'' One might be tempted to
criticize the generality of this definition, but it is that very
generality that makes it so appealing. For example, the complexity
envisioned here may be applied to all manner of natural and social
systems with the clear understanding that it is not objective, but
depends on both the system and the observer. In addition to the
subjective nature of this concept of complexity due to the explicit
inclusion of the observer in its definition, there is the additional
ingredient of subjectivity having to do with the ''questions'' the
observer asks of the system. Here the notion of observer is really
that of an experiment, but one could also formulate the same
definition involving systems containing human beings or other
conscious entities. It is difficult if not impossible, at the present
time, to know if conscious and unconscious matter can be described in
a unified way. Conscious matter, due to its complexity, can have
emergent properties which do not imply an animistic theory, but in
fact are consistent with a materialistic approach.

 The theory
of emergence is based on the idea that if we proceed towards
organization levels more complex there are new emerging properties not
present at the less complex levels (brain, consciousness, life are all
examples of this). The most important problem at this point is how to
consider the emergence of these new properties. They may be considered
as genuinely unknown or unknown for our temporary scientific
limits. The proponents of CAS would certainly argue that the existence
of emergent properties, which is to say self-organization, is partly
the reason for the development of their theory.  However, CAS is still
in its infancy and whether or not it will survive to adolescence, much
less to maturity remains to be seen.  
 One theory that has since its
inception included the effect of the observer
 on the system observed
is that of quantum mechanics (by using the case of
 quantum mechanics
we can appreciate again how much the observer is involved in a
physical 
 fact and in its analysis). 
 The paradigm of quantum
mechanics is useful because it is accepted as the most fundamental
picture
 of the interactions of matter in the universe and it too
includes this
 dichotomy of the system and observer. In this general
scheme the observer
 may also be included as part of the system and
therefore the observation may
 or may not change the system,
depending on the questions asked. For example,
 the observer might
influence the spectral lines in the light from alpha
 centuri, but
not the interest rates at the local bank.
 
 So now we come down to
the crucial question: Can we define a measure of
 complexity that
will be useful across a broad spectrum of problems, from the
 stock
market to superconductivity, from social discontent to the laughter
of
 children? Using Ruelle's definition one may think that it is
possible to
 define a measure, since the concept of ``difficulty''
admits an ordering
 relation. In other words it makes sense to say;
``problem A is more
 difficult than problem B''. However, a little
reflection reveals that this
 kind of relation is not really
objective, since the difficulty of the
 problem depends on whom has
to face it and at what time. Establishing
 objective criteria, for
example imagining that the problems have to be faced
 by machines
such as computers or by adopting rigid algorithms, just
 transfers
the subjectivity to the level of the arbitrary criteria adopted.
Herein we show that physical science, presumably our best effort at
constructing an objective theory of reality, is not immune from
these
 difficulties and we examine various proposals to overcome
them. Because of the 
 way we have based our discussion until now,
here we decide on a point where to start, 
 an assumption to use in
order to build up our system of ideas.

 We make the
assumption that complexity is a property of the system and we do
 not
address the difficulties associated with the observer, such as
prejudice, limited resources and so on. Even in this restricted
context of
 theory we hope that the measures discussed shall be of
some value.
 
\section{Complex Systems}

There has been a substantial body of mathematical analysis
regarding complexity and its measures and it is rather surprising the
broad range over which mathematical reasoning and modeling has been
applied. One class of problems that defines the limits of
applicability of such reasoning is denoted ``algorithmic complexity'.
A problem is said to be algorithmically complex if to compute the
solution one has to write a very long algorithm, essentially as long
as the solution itself. Applications of this quite formal theory can
be found in a variety of areas of applied mathematics, but even if one
restricts the discussion to the physical science, it is not possible
in a short space to give a fair description of the activity. Thus, we
focus our attention on nonlinear dynamics, system theory and
complexity in the physical sciences. Hopefully one shall be able to
extrapolate from this discussion to other areas of investigation. In
Section 6 we shall outline a vision of complexity whose connections
with algorithmic complexity became clear in the last few
years.

 A \emph{system} consists of a set of elements together
with a defining set of relations among those elements. All the
phenomena of interest to us here shall be viewed as a system. It is
also possible to study a subset of elements, called a \emph{subsystem}
of the system. Finally, the system may interact with the observer who
may be a member of the system itself or of the \emph{environment}. It
is also possible, and sometimes necessary, to define an environment of
the environment, and so on. As already pointed out, the complexity of
a system depends on the information sought by the observer, and this
depends on the purpose of the study. We imagine that a system may be
studied to ``understand it'', namely to describe and control it or to
predict its dynamics. For example, the weather cannot be controlled,
but it is very useful to make accurate short-term
forecasts. Predicting the trajectory of a hurricane may save millions
in dollars, not to mention the saving of lives, even if in principle
we cannot know its fundamental nature. It is often crucial to study,
whenever possible, the response of a complex system to external
perturbations. It is the set of these responses that constitute the
information that the observer tries to extract from the system and it
is the difficulty encountered in understanding, controlling or
predicting these responses that is intuitively used as the measure of
complexity. In the last part of this paper we shall outline a vision
of complexity, from which many papers of these Proceedings emerged,
which seems to be convenient to address practical issues of this kind,
in addition to shedding light onto the problems discussed in the first
part of this paper.

 It is useful to list the properties
associated with the complexity of a system, because we are seeking a
quantitative measure that may include an ordinal relation for
complexity. We note however that in everyday usage phenomena with
complicated and intricate features having both the characteristics of
randomness and order are called complex. Further, there is no
consensus among scientists, poets or philosophers on what constitutes
a good quantitative measure of complexity. For instance, Phil Winsor
in the abstract of his contribution to these Proceedings \cite{phil}
claims that his paper addresses philosophical and conceptual issues
departing from the usual mode of presentation of scientific journal
articles. Yet, we are convinced that his presentation reinforces the
perspective of complexity that we shall outline in Section 6, and that
this perspective would make it possible for us to express his views
with the scientific jargon of statistical mechanics, even if anomalous
statistical mechanics. 

For the time being, in this preliminary
exploratory phase, let us limit ourselves to remarking that any list
of traits of complexity is arbitrary and idiosyncratic, but given that
disclaimer the following traits are part of any characterization of
complexity \cite{weaver,flood}:

 i) A complex system shall
typically contain many elements. As the number of
 elements increases
so too does the complexity.
 
 ii) A complex system typically
contains a large number of relations among
 its elements. These
relations usually constitute the number of independent
 dynamical
equations that determine the evolution of the system.
 
 iii) The
relations among the elements are generally nonlinear in nature,
often being of a threshold or saturation character or more simply of
a
 coupled, deterministic, nonlinear dynamical form. The system often
uses
 these relations to evolve in a self-conscious way.
 
 iv) The
relations among the elements of the system are constrained by the
environment and often take the form of being externally driven or
having a
 time-dependent coupling. This coupling is a way for the
system to probe the
 environment and adapt its evolution for maximal
survival.
 
 v) A complex system typically remembers its evolution
for a long time and is
 therefore able to adapt its behavior to
changes in internal and
 environmental conditions.
 
 vi) A complex
system is typically a composite of order and randomness, but
 with
neither being dominant.
 
 vii) Complex systems often exhibit
scaling behavior over a wide range of
 time and/or length scales,
indicating that no one or few scales are able to
 characterize the
evolution of the system.
 
 These are among the most common
properties selected to characterize complex systems, see for example
\cite{SFI}, and in a set of dynamical equations, these properties can
often be theoretically kept under control by one or more
parameters. The values of these parameters can sometimes be taken as
measures for the complexity of the system. This way of proceeding is
however model-dependent and does not allow the comparison between the
complexities of distinctly different phenomena, or more precisely
between distinctly different models of phenomena. It is also worth
mentioning here that the recent results illustrated in Section 6 lead
us to add to this list a further trait:

viii) Complex systems
conflict with the stationary assumption and exhibit aging
properties. 

 In the above list we included one of the most subtle
concepts entering into
 our discussion of complexity, that is, the
existence and role of randomness 
 \cite{peter,arnold}. Randomness is
associated with our inability to predict
 the outcome of a process
such as the flipping of a coin or the rolling of a
 die. It also
applies to more complicated phenomena, for example, when we
 assume
we cannot know the outcome of an athletic contest such as a
basketball or football game, or more profoundly when we cannot say
with
 certainty what the outcome of a medical operation such as the
removal of a
 cancerous tumor will be. From one perspective the
unknowability of such
 events has to do with the large number of
elements in the system, so many in
 fact, that the behavior of the
system ceases to be predictable \cite{mackey}.
 On the other hand, we
now know that having only a few dynamical elements in
 the system
does not insure predictability or knowability. It has been
demonstrated that the irregular time series observed in such
disciplines as
 economics, chemical kinetics, physics, logic,
physiology, biology and on and
 on, are at least in part due to chaos
\cite{lasota}. Technically chaos is a
 sensitive dependence of the
solutions to a set of nonlinear, deterministic,
 dynamical equations
on initial conditions. Practically chaos means that the
 solutions to
such equations look erratic and may pass all the traditional
 tests
for randomness even though they are deterministic. Therefore, if we
think of random time series as complex, then the output of a chaotic
generator is complex. However, we know that something as simple as a
one-dimensional, quadratic map can generate a chaotic sequence. Thus,
using
 the traditional definition of complexity, it would appear that
chaos implies
 the generation of complexity from simplicity. This is
part of Poincar\'{e}'s
 legacy of paradox. Another part of that
legacy is the fact that chaos is a
 generic property of nonlinear
dynamical systems, which is to say chaos is
 ubiquitous; all systems
change over time, and because they are nonlinear,
 they manifest
chaotic behavior.
 
 A nonlinear system with only a few degrees of
freedom can generate random
 patterns and therefore has chaotic
solutions. So we encounter the same
 restrictions on our ability to
know and understand a system when there are
 only a few dynamical
elements as when there are a great many dynamical
 elements, but for
very different reasons. Let us refer to the former random
 process as
\emph{noise}, the unpredictable influence of the environment on
 the
system of interest. Here the environment is assumed to have an
infinite
 number of elements, all of which we do not know, but they
are coupled to the
 system of interest and perturb it in a random,
that is, unknown, way 
 \cite{zernike}. By way of contrast chaos is a
consequence of the nonlinear,
 deterministic interactions in an
isolated dynamical system, resulting in
 erratic behavior of at most
limited predictability. Chaos is an implicit
 property of a complex
system, whereas noise is a property of the environment
 in contact
with the system of interest. Chaos can therefore be controlled
 and
predicted over short time intervals whereas noise can neither be
predicted nor controlled except perhaps through the way it interacts
with
 the system.
 
 The above distinction between chaos and
control highlights one of the difficulties of formulating an
unambiguous measure of complexity. Since noise cannot be predicted or
controlled it might be viewed as being simple, thus, systems with many
degrees of freedom that manifest randomness may be considered
simple. This point requires an explanation. The literature on
stochastic processes shows that ordinary environmental noise, assumed
to be white, yields ordinary, and simple, diffusion equation, with the
same second derivative with respect to space as that appearing in the
ordinary Shr\"{o}dinger equation of quantum mechanics. In Section 6 we
shall mention conditions where the environmental fluctuations, being
correlated, cannot yield simple equations.  On the other hand, a
system with only a few dynamical elements, when it is chaotic, might
also be considered to be simple. In this way the idea of complexity is
again ill posed and a new approach to its definition is required.

In the earlier papers on systems theory it is argued that the increasing
complexity of an evolving system can reach a threshold where the system is
so complicated that it is impossible to follow the dynamics of the
individual elements, see for example, Weaver \cite{weaver}. At this point new
properties often emerge and the new organization undergoes a completely
different type of dynamics. The details of the interactions among the
individual elements are substantially less important than is the
``structure'', the geometrical pattern, of the new aggregate. This is the
self-aggregating behavior required in the CAS theory. Increasing further the
number of elements or alternatively the number of relations often leads to a
complete ''disorganization'' and the stochastic approach becomes a good
description of the system behavior. If randomness (noise) is to be
considered as something simple, as it is intuitively, one has to seek a
measure of complexity that decreases in magnitude in the limit of the system
having an infinite number of elements. We shall see that this attractive
goal is difficult to attain and all attempts to obtain and measure noise
rather than chaos either break the laws of physics or the \emph{ principle of
subjectivity}.

\section{Entropies}

Historically thermodynamics was the first discipline in physics to
systematically investigate the order and randomness of complex systems,
since it was here that the natural tendency of things to become disordered
was first observed. As remarked by Shr\"{o}dinger in his groundbreaking work
{\it What is Life}? \cite{life} : ``The non-physicist finds it hard to believe that really the ordinary laws
of physics, which he regards as prototype of inviolable precision, should be
based on the statistical tendency of matter to go over into disorder''.

In this context the quantitative measure of ``disorder'' that has proven to
be very valuable is \emph{ entropy} and the idea of thermodynamic equilibrium
is the state of maximum entropy. Of course, since entropy has been used as a
measure of disorder, it can also be used as a measure of complexity. If
living matter is considered to be among the most complex of systems, for
example the human brain, then it is useful to understand how the enigmatic
state of being alive is related to entropy. Shr\"{o}dinger maintained that a
living organism can only hold off the state of maximum entropy, that being
death, by absorbing negative entropy, or negentropy, from the environment.
He points out that the essential thing in metabolism is that the organism
succeeds in freeing itself from all the entropy it cannot help producing
while alive.

We associate complexity with disorder, which is to say with limited
knowability, and order with simplicity or absolute knowability. This rather
comfortable separation into the complex and the simple, or the knowable and
the unknowable, in the physical sciences will be shown to breakdown once a
rigorous definition of entropy is adopted and applied outside the restricted
domain of thermodynamics, in spite of Shr\"{o}dinger's best efforts. We shall
see that because of the fundamental ambiguity in the definition of
complexity, that even adopting the single concept of entropy as the measure
of complexity leads to multiple definitions of entropy some of which are in
conflict with one another.

We review the various definitions of entropy along with other various
measures of disorder that have been used in the physical sciences. The
definition of entropy as a measure of ''disorder'' encounters the same
problems of subjectivity that we found when we defined complexity. Order is
something that is difficult to define, and strictly speaking depends on the
questions that are asked of the system. The process of rendering objective
the concept of disorder leads to several quantitatively different
definitions of entropy. Once a definition is adopted everything is somewhat
''objective''. Unfortunately, as we already mentioned, the different
proposals lead to different quantitative results even though they are the
most objective measures of complexity we have available. The subjectivity
enters through the choice of definition that is the most suitable to answer
the questions of interest.

Following Cambel \cite{cambel} we roughly divide entropies into three
categories: the macroscopic, the statistical and the dynamical. In the first
group we find the entropy stemming from thermodynamics, for example the
original S-function entropy of Clausius as used by Boltzmann \cite{boltz} and
subsequently by Prigogine \cite{prigogine}. In the second category is placed
the entropy stemming from the assumption of a probability distribution to
characterize the system, such as the Gibbs entropy \cite{gibbs}. Here the
activity on the microscale, or to the dynamics of the individual elements in
phase space, is related to what occurs macroscopically, or at the system
level. A more recent formulation of the probability entropy is that of
Tsallis \cite{tsallis}. A special role in the statistical entropies is played
by the information entropy of Shannon and Kolmogorov, that relate the
physical properties of the system to the concept of information 
\cite{kolmogorov}. Finally, the dynamical entropies such as Kolmogorov's are
derived from the geometry of the systems dynamics in phase space. Other
possible choices for categories might include phenomenological,
informational or geometrical, but these would have no distinct advantage
over those above and would in part overlap with the categories chosen.

\subsection{Clausius' Entropy}

Entropy, like the length of a rod or the temperature in the room, is a
physical quantity that is measurable. At the absolute zero of temperature
the entropy of any piece of matter is zero. When the substance is brought
into any other state by slow, reversible little steps the entropy increases
by an amount that can be computed by calculating the ratio of the heat
supplied to the absolute temperature at which the heat was supplied and
adding up all these small contributions.

The \emph{ Second Law of Thermodynamics}, as formulated by Clausius in 1850,
states that it is not possible to conduct an experiment in an isolated
system whose only result is the spontaneous transfer of heat from a cold
region to a hot region, since if the system is isolated work cannot be done
on it. Consequently, this flow of heat defines a directionality (arrow) for
time. It then took Clausius fifteen more years to prove that the
thermodynamic temperature was an integrating factor for the quantity of heat
transferred, so he defined the function S (entropy) in the exact
differential form discussed above, meaning that the total entropy is
obtained by summing the above ratio of heat to temperature over any
reversible path of thermodynamic equilibrium states. Thus, this concept of
entropy implies that the system is macroscopic and isolated and requires the
existence of thermodynamic equilibrium. If these conditions are not
fulfilled then we cannot calculate the entropy exactly but must be satisfied
with the inequality $\Delta S\geq 0$ over a thermodynamic cycle, where the
equality only applies for a reversible process. The inequality means that
the change in entropy over a cycle, where $\Delta S$ is the difference in
the entropy at the beginning and the end of the cycle, is an increase for
irreversible processes or it is zero for reversible processes. The arrow of
time is therefore recovered as the direction of increasing entropy for
isolated systems and this in turn has been used to define what is actually
irreversible.

According to the thermodynamical entropy discussed here, a completely
random system would have maximum entropy and therefore maximum
complexity. This concept of entropy is very useful for studying large
scale physical and chemical systems at or near equilibrium. On the
other hand an ordered system, such as a living organism, would have a
low thermodynamical entropy and would therefore be simple under this
classification scheme. Since this conclusion runs counter to our
expectation that living systems are among the most complex in the
universe, we cannot simply apply the definition of thermodynamical
entropy to a system in order to determine its complexity. However, we
do not want to abandon the notion of entropy altogether since it is
tied up with the order-disorder property of a system. Thus, we shall
explore some of the extensions and refinements of the entropy concept
to see if these will serve our needs better. In section 6 we shall
conclude our search by noticing that the ordinary form of Gibbs
entropy plays a useful role, not so much to measure the complexity of
a state, but rather the condition of transition from dynamics to
thermodynamics.

\subsection{Boltzmann's Entropy}

The definition of entropy as it was introduced into thermodynamics by
Clausius did not rely on any statistical concepts. However in our
interpretation of order and disorder fluctuations certainly played a role.
It was the nineteenth century physicist Boltzmann that first attempted a
synthesis of the deterministic predictability of mechanics and
thermodynamics through studying the transport of large numbers of particles
in gases. He developed quite complicated equations that described the
fluid-like motion of gases including the collisions among the individual
particles. These collisions, reasoned Boltzmann \cite{boltz}, would produce
a randomization of the motion of the gas particles since one could not
determine with absolute precision the location and size of the individual
collision events. In this way he introduced a probability density function
that depended on the location and velocity of each of the particles in the
gas. His investigations lead him to introduce entropy in the form

\begin{equation}
\label{boltzmannprinciple}
entropy=k_B\log W\mbox{ .}  \label{ent1}
\end{equation}
Here $k_B${\it \ }is a constant that has the appropriate dimensions for
entropy and has come to be called Boltzmann's constant and the function {\it %
W} is a quantitative measure of the microscopic disorder in the system.

This very different looking form for the entropy, it is not Clausius' ratio
of heat to temperature, shares with the energy the property of 
{\it extensivity}, which means that if one considers 
two independent systems {\it A}$_{1}$and {\it A}$_{2}$ with entropies 
{\it S}$_{1\mbox{ }}$and {\it S}$_{2}$, respectively, then 
the entropy of the combined system is just the
arithmetical sum {\it S}$_{1\mbox{ }}$+ {\it S}$_{2}$, as it would for the
energy. The entropy is extensive through the logarithmic assumption which
means that the measure of disorder, {\it W,} for the combined system, {\it W}%
$_{com},$is given by the product of the individual{\it \ W}'s, i.e., {\it W}$%
_{com}=${\it W}$_{1}${\it W}$_{2}.$ The quantity {\it W} indicates disorder
that is in part due to heat motion in a system and in part due to the
different kinds of particles that can generally intermix in a
thermodynamical system. If we imagine that the phase space for an isolated
system can be partitioned into a large number of cells and that each cell is
statistically equivalent to each of the other cells, which is to say that
the probability of a particle occupying any of the cells in phase space is
equally likely, then {\it W} is the volume of phase space consistent with
the total energy of the system.

This definition of entropy given by (\ref{ent1}) is fairly abstract, depending
as it does on a volume element of the phase space for the microscopic
elements of the dynamical systems. Boltzmann also expressed the entropy in
more physical terms through the use of the continuous phase space
distribution function, $\rho \left( {\bf x},{\bf v},t\right) $, where {\bf x}
is the physical location of the all {\it N} particles in configuration space
${\bf x}=\left\{ {\bf x}_{1},{\bf x}_{2},...,{\bf x}_{N}\right\} $ and {\bf v} 
is the velocity vector of all {\it N} particles in velocity space ${\bf v}%
=\left\{ {\bf v}_{1},{\bf v}_{2},...,{\bf v}_{N}\right\} ,$ so this one
function keeps track of where all the particles in the system are as a
function of time and what they are doing. Boltzmann was then able to show
that the entropy could be defined in terms of the phase space distribution
function as

\begin{equation}
\label{eq2}
entropy=-k_B\int d{\bf x}d{\bf v}\rho \ln \rho  \label{ent2}
\end{equation}
which is a non-decreasing function of time. He was able to show that this
definition of entropy attains its maximum value when the system achieves
thermodynamic equilibrium, in complete agreement with Clausius' notion of
entropy.

We shall refer to the definition of entropy as given by Boltzmann as the
statistical entropy. This development reached maturity in the efforts of
Gibbs \cite{gibbs}, who was able to provide the mechanical basis of the
description of thermodynamic phenomena through the formulation of
statistical mechanics. Gibbs gave a probability interpretation to the phase
space distribution function, and introduced the notion of ensembles into the
interpretation of physical experiments.

The above statistical definition of entropy is very general and is similar
to the measure of complexity we seek. In fact if the system is simple and
thus we are able to measure all the coordinates and the momenta of all the
particles with extreme precision, we have from Eq.(\ref{eq2}) that this entropy
is a minimum. A simple system, namely one that is closed and perfectly
integrable will not have any growth of entropy, due to the
time-reversibility of the dynamical equations. Here ``integrable'' and
``time-reversible'' dynamics means that the particles obey Newton's laws of
motion. Even in the case where our measures are not infinitely precise, the
growth rate is small, as will become clear in a subsequent section when we
introduce the Kolmogorov-Sinai entropy. The probability definition of
entropy also has the advantage that it recovers Clausius' proposal in the
statistical limit. Unfortunately the assumption made by Boltzmann is not
easily checked, but, on the contrary the truth of Eq.(\ref{eq2}) would
contradict physical law and therefore in principle can not be true. This
impossibility has been proved in a variety of ways, but still 
{\it Boltzmann's dream} puts us on the path to cross a bridge from dynamics to
thermodynamics, from reversible, microscopic processes to irreversible,
macroscopic processes. The latter is what we know and can directly measure,
the former is a useful hypothesis that has been indirectly measured, but the
connection between the two remains a mystery.

\subsubsection{Prigogine's balance equation}

The second law of thermodynamics is so well grounded in experiment that it
provides a guide to every possible definition of entropy. Thus, we know that
whatever definition we choose, entropy must increase or remain constant in a
closed system, or more precisely in the thermodynamical limit (where the
system is described by an infinite number of variables) it must be a
non-decreasing function of time for a closed system. This regression to
equilibrium, where as we mentioned equilibrium is the most disordered
configuration of the system, obtained by Boltzmann is the irreversibility
property of the entropy identified by Clausius and defines the arrow of
time. The future is therefore the direction in time that decreases the
amount of order and leads towards the ``heat death'' of the universe. But as
we well know this does not occur equally at all places and at all times; not
all systems deteriorate in the short term, if they did then life would not
be possible. One way to quantify the local increase in the entropy of a
system was developed by Prigogine \cite{prigogine}.

Shr\"{o}dinger \cite{life} identified negentropy as that quantity a living
organism obtains from the environment in order to keep from dissipating,
that is the ``stuff'' that enables the organism to maintain its order.
Prigogine \cite{prigogine} was able to develop and generalize this concept
through the formation of \emph{dissipative structures} that are maintained
through a flux of material and/or energy through the system of interest.
Explicit in these ideas is that order is maintained by means of the system
interacting with the environment, which means that the system is not closed
as it was in the discussion of Clausius and Boltzmann. The dissipative
structures of Prigogine are maintained through fluctuations providing
sources of energy to the system from the environment and dissipation
extracting energy from the system to the environment. This balancing of
fluctuations and dissipation maintain the flux through the system which in
turn supports the organization of the dissipative structure. This balance of
mechanisms was expressed in terms of the changes in the statistical entropy
by Prigogine:

\begin{equation}
\Delta S_T=\Delta S_I+\Delta S_E  \label{ent3}
\end{equation}
where $\Delta S$ is the change in entropy, and the subscripts refer to the
internal entropy change ({\it I}), the entropy change in the environment (%
{\it E}) and the total entropy change ({\it T}) of the system. The arguments
of Clausius, Boltzmann and Gibbs apply to the internal entropy of the system
$\Delta S_I$ which is zero for a closed system. Thus, even though 
$\Delta S_I $ is non-decreasing over time, the change in entropy of the system 
$\Delta S_T$ can be positive, zero or negative, depending on the contribution
of the environment to the entropy change, $\Delta S_E$. The system can
extract what it needs from the environment to generate and/or maintain its
order. Thus, the ordering of the system is more than compensated by a
disordering of the environment. Consequently, as the knowability of the
system increases due to $\Delta S_T<0$ and the knowability of the
environment decreases due to $\Delta S_E<0$, indicating that the negentropy
extracted from the environment to enhance the order of the system increases
the disorder of the environment. Systems in which $\Delta S_T<0$ are said to
be self-organizing and must occur under conditions where the system is far
from equilibrium, otherwise the internal and environmental contributions to
the entropy change would just cancel one another.

It is worth remarking that the vision of complexity emerging from this
paper, as discussed in section 6, does not rule out the Prigogine
perspective as a source of pattern formation. However, it makes an
additional possibility emerge, which is not out of equilibrium
thermodynamics. As we shall see, it is a condition intermediate
between dynamics and termodynamics. 

\subsection{Shannon and Kolmogorov-Sinai entropy}

As we have seen, the thermodynamical entropy can be given a dynamical
interpretation, but to do so one has to interpret the dynamics using a
probability density. This procedure is questionable and has given rise to
paradoxes and controversies that remain unresolved. A rigorous mathematical
treatment, based either on quantum or on classical mechanics of closed
(independent) systems, does not produce any regression to equilibrium, to
the contrary it results in an eventual return to the initial state of the
system after waiting a sufficiently long time. This is another of those
paradoxes of Poincar\'{e}, a dynamical system following Newton's laws will
return to its initial state infinitely often over time. This recurrence
property of Poincar\'{e} rules out the possibility of a dynamical system
relaxing to an equilibrium state using only the equations of motion.

The recurrence of dynamical systems has necessitated the introduction of a
number of hypotheses to account for the arrow of time which is so evident on
the scale of biological evolution. To explain the relaxation of a system to
equilibrium and therefore to give time its direction, physicists allow for a
certain uncertainty in the measured values of dynamical variables which is
referred to as coarse-graining. This is traditionally done by discarding the
fiction of a closed system and recognizing that every system has an
environment with which it interacts. By explicitly eliminating the
environmental variables from the description of the system dynamics one
obtains a description that is statistical in nature. The absolute
predictability which was apparently present in the deterministic nature of
Newton's equations is abandoned for a more tractable description of the
system having many fewer variables, but the price has been high and that
price is predictability. Since we cannot know the environment in a finite
experiment, the environment is not under our control, only the experimental
system is accessible to us, the environments influence on the experimental
system is unpredictable and unknowable except in an average sense when the
experiment is repeated again and again. It is this repeatability of the
experiment that allows us to map out all the different ways the environment
influences the system through the construction of the \emph{ ensemble
distribution function} that captures all the available information, features
common to all the experiments in the ensemble.

Unfortunately there does not exist a systematic way to include all the ways
the environment can interact with an arbitrary dynamical system and so the
desirable situation outlined above has not as yet been attained. One way
that this program has been pursued has been to make the evolution of the
system uncertain by inserting a random force into the equations of motion.
This procedure saves the applicability of physical theories, but it also
introduces the seed of \emph{subjectivity} into the evolution of the system.
We have to develop mathematical tricks to treat systems with an infinite
number of degrees of freedom like the environment, where the global
properties of the overwhelming majority of degrees of freedom are chosen in
a subjective manner, mostly for computational convenience. One such trick is
to assume that the environment is already in the state of equilibrium before
interacting with the system, and that it is so large that it remains in
equilibrium throughout its interaction. Therefore it is assumed that one
knows nothing of the environment at the start of the experiment and that one
can learn nothing about the environment from the results of the experiment.
This is a sad commentary on the present state of statistical physics and
what it can teach us about complex systems.

We anticipate what will become clear shortly, namely that the problem with
existing dynamical theories in so far as they are inconsistent with the
statistical interpretation of entropy is that they are deterministic and
time-reversible. These two properties imply that no probabilistic treatment
of dynamics is objective, and the correctness of a statistical picture stems
from subjective assumptions. Therefore there is no direct connection between
dynamics and thermodynamics, because the connecting link, that being
statistical mechanics, requires the introduction of probability theory. In
the same vein, the \emph{ Correspondence Principle}, the principle according
to which quantum mechanics and classical mechanics are made equivalent, is
always used in a statistical sense, which without subjective assumptions
cannot be correct.

Implicit in the concept of entropy is the idea of uncertainty. The latter
idea only makes sense in a context where there is a conscious being that is
extracting information from the system, and is therefore subjective.
Uncertainty means that not all the information one needs for a complete
description of the behavior of a system is available. Even the term
``needs'' is in this sense subjective, because it depends on the questions
the observer poses, which in turn depends on the ''purpose'' of the
observer. This is where all subjectivity enters, and we do not go further
into the philosophical implications of having an observer with a purpose
conducting the experiment. We only wish to be clear that a system containing
conscious individuals can not be treated in a deterministic way since the
objectivity stemming from determinism conflicts with the subjectivity of the
individuals (free will).

However, we can safely say that entropy is a measure of uncertainty, and
like uncertainty, entropy is a non-decreasing function of the amount of
information available to the observer. This connection between information
and thermodynamics is quite important, and at this stage of our discussion
we can say that it is the uncertainty that allow us to describe dynamical
systems in thermodynamical terms.

Shannon \cite{shannon} determined how to construct a formal measure of the
amount of information within a system and the problems associated with the
transmission of a message within a system and between systems. He expressed
that information in terms of \emph{ bits}, or the number of binary digits in a
sequence. He was able to prove that a system with {\it N} possible outputs,
where the output {\it i} had the probability of occurring $p_i,$ can be
described by a function {\it H} that attains its maximum value when each of
the possible states of the system have the same probability of occurrence,
that is the assumption of maximal randomness (maximum uncertainty) in which
case $p_i=1/N$. This result is essentially equivalent to Gibbs' treatment of
Boltzmann's entropy, where the function {\it H} is equivalent to
Shr\"{o}dinger's negentropy. The analytic expressions for the entropy are
exactly the same, but this informational interpretation offers possibilities
of extending the definition of entropy to situations using conditional
probabilities, resulting in conditional entropies, mutual entropies, and so
on. This means that it is possible to recognize two equivalent pieces of
information, and to disregard the ''copy'' because nothing new is learned
from it. It is possible to extract the new pieces of information from a
message of which the majority of the content is already known, and therefore
it is useful for separating the knowable from the known.

New pieces of information decrease the level of uncertainty, and thereby
increase the order of a system. As mentioned above this is precisely the
mechanism discussed by Shr\"{o}dinger using his concept of negentropy. This
fact is highlighted by the famous paradox of Maxwell's demon. The demon
provides a mechanism by means of which a closed thermodynamic system could
decrease its entropy by using information within the system itself in
apparent violation of the Second Law of Thermodynamics. The paradox was
finally resolved by the physicist Leo Szilard \cite{szilard} who calculated
the entropy associated with the demon's acquisition and use of information
and found that it exactly equalled the amount of entropy by which the system
was reduced due to the demons efforts. Thus, the demon generates as much
entropy as she suppresses and there is no violation of the second law.

The information entropy is closely related to the Kolmogorov-Sinai
(KS) entropy. We have already quoted the relevant work of Kolmogorov
\cite{kolmogorov}. The contribution of Sinai to this entropy is given
by the work of Ref. \cite{sinai}. We do not address the delicate
mathematical concepts behind this important kind of entropy. We limit
ourselves to pointing out that the KS entropy is a trajectory
property, made computable, in the case of the dynamical chaos by the
Pesin theorem \cite{pesin}, which establishes this entropy to be given
by the sum of the positive Lyapunov coefficients.  How to relate this
trajectory property to the entropy of Eq. (\ref{eq2}), which is, an
ensemble property?  This important issue has been discussed by Latora
and Baranger \cite{latorabaranger}. The very interesting work of these
two authors rests on the assumption that the density time evolution can
be reproduced by a bunch of trajectories, which, due to the fact that the
Lyapunov coefficients are finite, tend to spread, thereby occupying an
increasing number of cells of the phase space. We note that according
to Petrosky and Prigogine\cite{tomio1,tomio2}
an equation of time
evolution for densities, once defined, must be considered as a law of
physics on its own. Consequently, it is a problem of some interest to
establish a connection between the entropy of Eq. (\ref{eq2}) and the
KS entropy without invoking the trajectory time evolution. This
delicate problem has been studied by the authors of
Ref.\cite{markos}. These authors pointed out that in the case of
conservative systems the entropy of Eq. (\ref{eq2}) increases as an
effect of a coarse graining.  They also noticed that in the case of
intermittent randomness, even if the ergodic condition is assumed, so
as to properly define the KS entropy, it is impossible to make the
entropy increase of Eq. (\ref{eq2}) compatible with the KS entropy.
This is the first example of the problems caused by anomalous
statistical mechanics.

\subsection{R\'{e}nyi-Tsallis Entropy}
In this section we mention briefly that in the last few years there has been a great interest
for the R\'{e}nyi and Tsallis entropies. 
The hungarian mathematician Alfr\'{e}d  R\'{e}nyi \cite{renyi} in his treatise on
probability theory, has shown that one can actually build up ''information
functions'' that share the order relation property with Shannon's
information entropy (and therefore the metric entropy). When the set of
probabilities $\left\{ p_i\right\} $ are such that $\sum p_i\log _2p_i$
diverges, it is possible to find a real number {\it q} 
(0{<}{\it q}{<}1) such that

\begin{equation}
I_q\equiv \frac 1{q-1}\log _2\left[ \sum\limits_{i=1}^Np_i^q\right]
\label{ent4}
\end{equation}
converges. This is defined as the information function of order {\it q}.
Such information functions are useful when, going to the continuum, the
probability density $p\left( x,v;t\right) $ has long tails with diverging
moments. Such distributions are quite common in the social and life
sciences, and are found to be more prevelent in the physical sciences than
was once believed, see for example West and Deering \cite{west}.

Recently Tsallis \cite{tsallis} adopted a form for
the entropy, which, apparently, looks similar to the R\'{e}nyi form, namely
\begin{equation}
\label{onpurpose}
S_q\equiv \frac{1-\sum\limits_{i=1}^Np_i^q}{q-1}  \label{ent5}.
\end{equation}
Actually, this entropy violates the extensive nature of the Boltzmann
entropy, discussed in Section 3.2, and, consequently departs from the
R\'{e}nyi entropy, which is still additive.  The reasons of the
success of Tsalliis' entropy is that by maximization under suitable
constraints, it leads to equilibrium distributions with an inverse
power law form. This is an interesting property, even though it raises
the obvious criticism that this entropy is given its form, namely the
form of Eq. (\ref{onpurpose}), on purpose. This means that the
deviations from the standard equilibrium distribution are well known,
and are a consequence of the renormalization group approach
\cite{westtheseproceedings} and that the entropy of
Eq. (\ref{onpurpose}) is given its form for the specific purpose of
yielding an inverse power law distribution.

A more satisfactory approach, in our opinion, is the derivation of the
entropy form directly from dynamics. The work of Ref. \cite{marco}
proves that this is possible. An oscillator of interest playing the
role of thermometer is coupled to a dynamical system, called booster
for the specific purpose of keeping it distinct from the ordinary
thermal baths, which already rest on the assumption that
thermodynamics holds true. The authors of Ref.\cite{marco} aimed at
reaching their conclusions with no thermodynamic assumption
whatsoever. They built up a Fokker-Planck equation for the oscillator
of interest and used the width of the velocity distribution, expressed
in terms of dynamical properties, to measure the temperature of the
booster. The interesting result of this paper is that the Boltzmann
principle of Eq. (\ref{boltzmannprinciple}) is recovered from dynamics
in the limiting case of a booster with a large number of degrees of
freedom.  Note that the thermometer interacts with only one particle
of the booster, called a doorway particle. For this procedure to reach
the wished result, it is essential that the correlation function of
the coordinate of the doorway particle to undergo a relaxation process
fast enough.

It is important to point out that significant attempts
at applying the same procedure in the specific case where the
correlation function of the coordinate of the doorway particle
undergoes an inverse power law decay \cite{mario,massi,jacopo} have
been done. The original project of deriving out of this procedure the
Tsallis entropy \cite{anna} did not yield satisfactory conclusions. A
new and unexpected result emerged from these attempts. This has to do
with the fact that in the ordinary case of normal statistical
mechanics the transition from dynamics to thermodynamics is virtually
instantaneous. We make the conjecture that the process of transition
to the scaling regime in a diffusion process is an indicator of the
transition from dynamics to thermodynamics. In fact, the main
difference with the attainment of the ordinary form of equilibrium is
that there is no feedback on the bath, the diffusion caused by the
bath fluctuations being the only active process. With this perspective
in mind, the main result of the work of Ref. \cite{jacopo} is that the
process of transition from dynamics to thermodynamics lasts forever,
thereby leading us to consider this condition as a new state of
matter. The indication of this kind of complexity is not given by an
entropy measure, but it is disclosed by the detection of multi-scaling
properties.  We shall be referring to this state of matter as the
Living State of Matter (LSM). We shall discuss again the relevance of
this perspective for complexity in Section 6. Note that the work of
Allegrini et al \cite{allegrini} can be regarded as a pioneering
attempt in this new direction. 
 
\section{Does Physics Really Describe Reality?}

\subsection{Objectiveness and reductionism}

We have seen in the previous chapter how the concept of entropy has
developed into a useful tool for the study of complex phenomena. The
understanding of this development is a guide in our search for a
suitable measure for complexity. We learned that asking the wrong
question leads one along a false trail, however, posing the right
question enables one to gather information regarding the process of
interest. A collection of experiments can be used to quantify a
measure, and different methods of analysis provide various insights
into the properties of the phenomenon using that measure. We have also
seen that the concept of information offers the possibility of
treating every problem with a great deal of generality, but
nevertheless this approach runs the risk of confusing the physical
phenomena, with any simulation of them, since they share the same
amount of information. Subjectivity enters into this discussion in the
selection of the physical or mathematical property we choose to
investigate, and therefore into ``focusing'' (paying attention to) the
flow of information in order to increase the rate of data
processing.

 The fact that we repeat the experiments and that we
try new methods of
 analysis is important in order to see if our
description of the reality
 is well approximated. Nevertheless we
should notice that this does not
 mean that in this way we can
suppress any subjectivity. In fact we
 should take into account at
least two factors:
\begin{enumerate}
\item It is still the scientist who manipulates and interacts with the physical system.

\item The fact that we obtain many times the same results may be
related to the fact that we use all the time the same methods of
analysis, but these methods could be wrong for whatever unknown
reason.

\end{enumerate}

If we believe that these two objections are true, we should probably
look for physical power of prediction elsewhere.  we may reach a
satisfying compromise when we test not only nature but also our
methods of analysis.  If, however, the laws stemming from the
investigation depend on subjective limitations or other properties of
the observer (any observer belonging to the human species, for
instance, may be assumed to behave classically), then one may wonder
if some ``other observer'' with different constraints would see
different laws of nature. It may look strange that scientists are
involved in this kind of philosophical discussion, since it may appear
that a uncorrectable subjective statement or theory cannot be
scientific. However, this is not true. No discipline is immune from
this paradoxical logic. In fact the foundation of all of science,
which by consensus is physics, has this problem at its very core:
quantum mechanics.

A familiar example of the paradoxes in
quantum mechanics should suffice; let us consider Shr\"{o}dinger's
cat. Recall that the cat is in a box and cannot be observed. There is
poison gas that can be released by means of particle decay from a
radioactive sample that is also within the box. Since particle decay
is a quantum process, it is described by the superposition of a wave
function in which the particle has decayed and one in which it has not
decayed. Thus, determining whether the cat is alive or dead at any
particular time is ambiguous since it would appear that the state of
being of the cat is a superposition of a state in which the gas has
been released and the cat is dead and a state where the gas has not
been released and the cat is alive. Thus, there is a sense in which
the cat is both alive and dead at the same time. It is not true that
the quantum properties violate Aristotelian law of
non-contradiction. The thing is that logic, as physics, represents
different levels of approximation. An architect does not need Einstein
's relativity to build a house but just mechanics and statics. For
this reason his/her world would be well represented by classical
logic. We are not correct if we say that classical mechanics is not
true anymore, after that quantum theory has been formulated.  We are
incorrect exactly in the same way if we say that classical logic has
been invalidated by fuzzy logic. Every logic describes a different
level of reality, or if we prefer, the same reality from different
points of view.

 These quantum properties violating the normal
Aristotelian logic of the ``excluded third'' are completely objective
and experimentally verified at the atomic level. However, the theory
does not contain any parameter that is not observable at the level of
the cat. At a certain scale size the superposition of distinct states
is broken, and there is a random collapse of the total wave function
onto one of the two states, thereby implying that the cat is either
dead or alive but not both. This issue of measurement encapsulated in
the phrase, \emph{ collapse of the wave function}, remains one of the
mysteries of quantum mechanics. We still do not know the when and the
why of wave function collapse, but it is precisely at this unknown
level that the system ceases to be described in a deterministic way
and a probabilistic approach becomes necessary. It should be
emphasized that after the wave function collapse the uncertainty
regarding the health of the cat increases, since we have passed from a
deterministic picture to a statistical one, and therefore the entropy
increases as well. The good news is that the measurement is
irreversible, so that the arrow of time is recovered, but at this
point one may argue whether or not the observer has obtained any
information from the experiment. Do we have more information, or do we
have more uncertainty and therefore more entropy?

 The answer
to this question may seem unsatisfactory, because in its present
formulation Quantum Theory can not really be applied to the above
problem,
 because it does not apply to individual systems, only to
ensembles of
 systems. This means that either there are infinitely
many observers each
 conducting the same experiment, or there is one
observer conducting an
 infinite number of identical
experiments. Either of these two perspectives
 enables us to resolve
the paradox regarding the cat. The outcome of the
 experiment is
uncertain so that the entropy has increased, this situation
 arises
because for each of the experiments the system was prepared in
exactly the same way, in the state of minimal entropy. The situation
is
 different in the case of a single system, assuming that we only
have one
 cat. If quantum theory is applied to this case it should
tell us how and
 when the wave function collapses. We also need to
know if the measurement
 apparatus and the observer should be
included in the wave function. Even if
 a probabilistic approach is
adopted from the very beginning, it follows that
 infinitely many
identical systems depart from one another with alternative
 stories
on the fate of the cat. Do we then have to assume alternative
stories in infinitely many universes \cite{dewitt}? If we assume that
the
 observer and the macroscopic apparatus are classical, and
therefore obey the
 logical principle of the ``excluded third'', then
these paradoxes are
 resolved, but a unified theory is still
missing.

 This brief excursion into quantum theory
should be sufficient to show the
 unsatisfactory state of the
physical theory of measurement, since it cannot
 explain statistical
properties like entropy increase without encountering
 difficulties
with the principle of objectivity. For this reason we shall
subsequently return to an extended discussion of the influence of
quantum
 uncertainty on macroscopic knowability.
 
 An issue
related to the information paradigm of physical understanding of
nature is the \emph{principle of reductionism}. This principle, in a
nutshell, states that the process of understanding implies processing
data
 for the purpose of arriving at generalizations. Such
generalization are very
 efficient descriptions of the world,
reducing what we need to remember and
 enhancing our ability to
communicate with one another. It is much simpler to
 communicate a
law than it is to communicate the results of thousands of
experimental upon which the law is based. However, in its strong
forms, reductionism
 states that to understand complex phenomena one
needs only to understand the
 microscopic laws governing all the
elements of the system that make up the
 phenomena. This reasoning
implies that once one understands all the parts of
 a problem, one
can ``add them up'' to understand the total. The whole is
 just the
sum of its parts. That may sound fine in geometry but it is an
incomplete description of natural phenomena. The counterpoint to
reductionism
 is System Theory, that states that a system very often
organizes itself into
 patterns that cannot be understood in terms of
the laws governing the single
 elements. This self-organization
constitutes the \emph{ emergence} of new
 properties, that arise, for
example, in phase transitions. Living beings,
 too, cannot be
understood using reductionism alone, but a more wholistic
perspective has to be adopted. This change in perspective, from the
reductionistic to the wholistic, in some ways resembles the passage
from
 deterministic to probabilitistic knowledge. In both cases the
meaning of
 ``knowledge'' changes with the changing perspective. From
our arguments
 regarding physical theory we know that a complex
macroscopic system can be
 known in a reductionistic way in
principle, but not in practice, while at
 the same time it can be
known in a thermodynamical (wholistic) sense in
 practice, but not in
principle.

\subsection{Information, incompleteness and uncomputability}

In addition to the arguments given above, there might also exist other
reasons why, given our present state of knowledge, physical theories do not
provide a satisfactory description of reality. It is not sufficient for
physics to describe the world within the laboratory, it must also faithfully
describe the world in which we live. It seems clear that reductionism is not
enough to describe systems where the pattern of information flow often plays
a role more important than that of microscopic dynamics, for example, in
phase transitions. However, we still want the macroscopic rules to be
consistent with the microscopic ones. If new properties emerge, even if it
is impossible in practice to predict them from microscopic dynamics, they
must be implicit in the microscopic equations. This weak reductionistic
assumption is part of the objectivity principle. A chemical reaction is too
difficult to be explained from a purely quantum mechanical perspective at
the present time, but nevertheless no violation of quantum mechanics is
expected to take place during a chemical reaction. The analogous situation
arises at a higher level for the biological properties of a cell that cannot
be understood in terms of chemistry alone. Our understanding of the global
properties is achieved from a wholistic point of view, but the emerging
properties have to be compatible with a weakly reductionistic perspective.
Otherwise we would be tempted to imagine different laws for different space
and/or time scales, or different levels of complexity. This, in turn,
inhibits any possible mechanistic (objective) view of reality. We stress
that in our perspective the principle of objectivity, namely the objective
existence of mechanical laws, does not necessarily mean that the laws are
deterministic, but a seed of randomness may be involved. Actually we shall
argue that a seed of randomness must be involved in any fundamental
description of reality.

We stress again that there is no randomness involved in the classical
perspective, while in quantum systems randomness is triggered at the
level of measurement and is ultimately the cause of all the
paradoxes. Further, classical mechanics is time-reversible, and in the
absence of measurement so is quantum mechanics, thus, it is therefore
impossible to recover the arrow of time. The English physicist
R. Penrose \cite{penrose}, stressed in a recent book, another way in
which standard physical theories fail to describe reality. He
(Penrose) developed an extended argument devoted to rule out the
possibility of creating an artificial intelligence using standard
computers. In his discussion he explains how physics is basically
''computable'', which is to say that the laws of physics can be
faithfully implemented using computer programs, and cannot therefore
explain cognitive activity. Many scientists argue that awareness and
consciousness require properties that computers lack, see for example
\cite{lowenhard}. Penrose, however, proves that mathematical reasoning
is not computable. But Penrose himself judges these (human) properties
as a quality. No rules can substitute any human intuition. Namely, it
is impossible for any computer to have particular mathematical
knowledge available to our brain. The proof of this assertion requires
one to define what a computer is meant to be, or what is called in
mathematical jargon a ``universal Turing machine'', and what it can or
cannot do, even with unlimited time and available memory. Given these
constraints it is possible to use a version of the famous
incompleteness theorem of G\"{o}del \cite{goedel}, namely, that every
set of formal mathematical rules is always incomplete. In particular
the knowledge itself of this incompleteness is not available to formal
theories, but to us as human beings, and that is so because we are
able to understand the nature of ''paradox''. It has been proved that
the formal theories can be expressed in terms of computation and
vice-versa, so that our capabilities for going beyond what is
prescribed for formal theories by G\"{o}del's theorem is a conceptual
proof of the existence of non-computable phenomena in the
world.

As earlier stated, a natural application of computation
theory has been to the development of a measure of complexity. This
measure
 can be viewed as a generalization of Shannon's information
entropy. It is
 called ``algorithmic complexity'' or
Kolmogorov-Chaitin complexity \cite{kolmogorov,chaitin}, 
 after the
names of the two mathematicians that
 independently defined it. This
measure applies to binary strings and is
 defined as the length of
the string in bits for the shortest program that is
 able to compute
the string. Just like entropy, this function reaches a
 maximum if
complete randomness occurs, since genuine randomness is
non-computable, one has to specify the entire sequence in the
program. The
 \emph{Kolmogorov-Chaitin entropy}, like informational
entropy, enables one to
 define conditional or mutual properties, to
establish subadditive
 properties, that are the common features of
complex phenomena. This measure
 is very useful from a conceptual
point of view, but it does not have a
 practical use, since theorems
indicate that {\em it cannot be computed}. This
 particular
definition of entropy has been used as a measure of complexity in
 a
number of different fields, including program optimization as well
as
 image and information compression, but it is not useful for us
here.

\section{Randomness and determinism in physics}

\subsection{Reductionism and the end of physics}

We have argued that what has come to be called the science of
complexity is
 an interdisciplinary approach to the study of reality,
not confined to
 physics, but ranging from biology to economics, and
from there to
 psychology, neurophysiology and the study of brain
function, see for example
 Penrose\cite{penrose}. Schweber, in a
recent paper in Physics Today \cite
 {physics today}, pointed out a
crisis generated in physics by the success of
 renormalization group
theory: "The ideas of symmetry breaking, the renormalization group and
decoupling
 suggest a picture of the physical world that is
hierarchically layered into
 quasiautonomous domains, with the
ontology and dynamics of each layer
 essentially quasistable and
virtually immune to whatever happens in other
 layers. At the height
of its success, the privileged standing of high-energy
 physics and
the reductionism that permeated the field were attacked."
Reductionism was vigorously attacked early on by Anderson
\cite{anderson}, and he is not the only scientist who believed
reductionism had outlasted its usefulness.  
 The renormalization
group specifies a set of rules for establishing the
 critical
coefficients of phase transition phenomena. Wilson and Kogut \cite
{wilson} proved that the value of these coefficients can be assessed
with
 theoretical models in a way that is totally independent of the
detailed
 nature of elementary interactions. In other words, the
renormalization group
 approach establishes the independence of the
different levels of reality,
 and, even if in principle a man is
nothing more than a collection of atoms,
 his behavior has to be
studied, if ever possible, with scientific paradigms
 which do not
have anything to do with atomic dynamics. The leading role of
 high
energy physics in science was based on the implicit assumption that,
once the fundamental laws of physics are established, all phenomena,
at
 least in principle, can be explained. The advent of
renormalization group
 theory implies that even if a final theory is
possible, such as envisioned
 by Weinberg \cite{weinberg}, it cannot
be used to address the problems
 associated with the problems of
quantifying complexity. On the other hand,
 this dream of a final
theory might also be perceived as a nightmare by
 people like the
present authors, who hope and believe that reality is an
inexhaustible source of wonders. We share, on this issue, the same
view as
 Leggett \cite{legget}. We believe that the notion of strict
determinism must
 be abandoned and that the settlement of the problem
of the great unification
 in physics, even if it occurs, does not
represent the end of physics. At the
 end of his book Leggett
\cite{legget} concludes:"If even a small part of the above speculation
is right, then, far from the
 end of the road being in sight, we are
still, after three hundred years,
 only at the beginning of a long
journey along a path whose twists and turns
 promise to reveal vistas
which at the present are beyond our wildest
 imagination. Personally,
I see this as not pessimistic, but a highly
 optimistic,
conclusion. In intellectual endeavour, if nowhere else, it is
 surely
better to travel hopefully than to arrive, and I would like to think
that the generation of students now embarking on a career in physics,
and
 their children and their children's children, will grapple with
questions at
 least as intriguing and fundamental as those which
fascinate us
 today-questions which, in all probability, their
twentieth-century
 predecessors did not even have the language to
pose."

\subsection{White noise as a physical source for the fulfillment of the
Correspondence Principle}

As Mark Twain once remarked; ``The news of my death has been greatly
exaggerated.'' The same is true of the claims made regarding the
demise the
 discipline of physics; that its replacement by the
science of complexity may
 be premature especially if the paradigms
necessary to understand complex
 phenomena have their basis in
physical systems. It appears that the new
 paradigm upon which our
understanding of complex phenomena is based is that
 of randomness as
the key property of reality. This has presented a problem
 for modern
physics because of the conflict between the deterministic nature
 of
current theories and the consequent subjective character of randomness
as
 derived from these theories as we have discussed. Weinberg
\cite{weinberg}
 argues that quantum mechanics cannot be changed, and
that any possible
 generalization of this theory might prevent us
from keeping intact the whole
 corpus of facts that this theory
explains with such striking accuracy. On
 the other hand, we think
that quantum mechanics can be generalized, or, and
 this is probably
a more accurate perspective, that quantum mechanics can be
 recovered
from a new physical principle where randomness is held to be a
genuine property of nature. This can be done if the postulate of
measurement
 in quantum mechanics, wave function collapse, is
replaced by a dynamical
 ingredient which is genuinely stochastic
such as proposed by Ghirardi,
 Rimini and Weber
\cite{GRW}. Giovannetti et al. \cite{giovannetti} have
 shown that
rather than destroying quantum mechanics altogether, a concern of
Weinberg \cite{weinberg}, the addition of weak stochatic forces in
the
 microscopic domain results in physical effects difficult to
detect with
 current technologies, which would account for its not
being detected to
 date. However, such random forces would
legitimatize the assumptions
 invariably made so far in deriving a
unified picture of classical mechanics
 and thermodynamics from
quantum mechanics. This strategy has been adopted by
 Giovannetti et
al. \cite{giovannetti} and we are convinced that this
 paradigm of
randomness as reality rather than its being a consequence of
uncertainty or limited knowledge of initial conditions, provides the
proper
 perspective to discuss the problems of complexity. In part
because it
 implies that there is a fundamental limitation to what we
can know with
 absolute certainty about the nature of reality.
 
Since we do not live on the microscopic level, at least consciously,
it is
 of driving interest how to connect the macroscopic world of
our everyday
 experience, characterized by the existence of
thermodynamical processes, to
 the microscopic world of quantum
mechanics. This problem is subtlety related
 to that of deriving
classical mechanics from quantum mechanics. This is
 especially true
now because of the widespread conviction that chaos might
 enable us
to realize Boltzmann's dream of constructing a mechanical basis
 for
microscopic processes, see for example Lebowitz \cite{lebowitz}.
However, Boltzmann's dream is as elusive as the holy grail, and
establishing
 a direct manifestation of classical chaos in quantum
physics has so far not
 occured despite the efforts of thousands of
researchers, see the discussion
 by Reichl \cite{linda}.
 
 It
might be argued that since the Correspondence Principle shows us how
to
 make the transition from the quantum to the classical domans,
that it should
 also be able to guide us to the proper interpretation
of non-integrable
 systems in quantum mechanics. On the other hand,
classical physics can be
 portrayed as the physics of events, and the
paradoxes of quantum mechanics
 arise because of the lack of events
in the quantum domain, as described by
 Blanchard and Jadczyk
\cite{BJ}. The search for a unified picture of quantum
 and classical
mechanics, and therefore of the macroscopic and microscopic
 worlds,
has been undertaken by countless scientists. Many proposals for a
unified theory have been put forth, and some have direct bearing on
our own
 investigation into measures of complexity, but none have as
yet emerged as
 clearly superior to the others.
 
 According to
quantum mechanics the dynamics of a body is described by the
Shr\"{o}dinger equation, and the predictions of this fundamental
equation must
 be consistent with those provided by Newton's
equations. A microscopic body
 is predicted by quantum mechanics to
be characterized by a wave function
 evolution and the average
evolution must mimic the time evolution of a
 classical trajectory as
the mass of the particle becomes macroscopic. Thus,
 the chaos of
classical physics presents a problem for quantum mechanics in
 that
the center of gravity of the evolution of a wave function must
become
 erratic due to the chaotic evolution of the corresponding
classical
 trajectory of a particle. This is actually still an
unsolved problem, and
 the source of the difficulty may in fact be
due to the physics of chaos
 being strongly dependent on the
dynamical structures in classical phase
 space being fractal and
self-similar, see for example Mandelbrodt \cite
 {benoit} for a more
complete discussion. The fractal paradigm implies that
 the phase
space structures, when examined on any scale, look the same. There
is structure within structure within structure, down to the smallest
space-time intervals. This picture conflicts with quantum mechanics,
where
 there is a natural scale limitation imposed on the quantum
description by
 the \emph{ uncertainty principle}.
 
 Consider a
small volume of phase space in which an ensemble of initial
trajectories (classical particles) are placed, released and allowed
to
 evolve according to Liouville's Theorem (Newton's Laws). The
existence of
 chaotic solutions imples that the initial volume will
fragment, forming
 whorls and tendrils that interpenetrate all the
available phase space
 without changing its volume, forming an
interwoven fractal structure. The
 quantum picture of the same
process, on the other hand, requires that the
 process of
fragmentation cannot continue indefinitely. When the
 fragmentation
of the wave function reaches scales on the order of Planck's
constant, it is stopped and self-similarity is broken. In the last
decade
 scientist have wondered if this inhibiting effect of quantum
mechanics might
 manifest itself in dynamical processes otherwise
expected to be classical.
 
 Berry \cite{berry} argues that in the
case where quantum mechanics is expected to recover classically
chaotic trajectories, the exact correspondence between the time
evolution of a single wave function and a Newtonian trajectory is lost
on a time scale depending logarithmically on Planck's constant. This
time scale would make this effect, the in-equivalence between
classical and quantum phenomena, experimentally accessible. In spite
of this remarkable prediction, no significant effect has been found so
far. According to Roncaglia et al. \cite{roberto} this failure is, at
least in part, due to the fact that the comparison between quantum and
classical predictions must be made at the statistical level, within
the so-called Gibbs ensemble perspective that we discussed
earlier. Roncaglia et al. \cite{roberto} further argue that there
might be discrepancies between quantum mechanical and classical
prescriptions at the statistical level if the experimental observation
of the influence of chaos moves from the case of ordinary to that of
anomalous diffusion. In Section 6 we shall come back to this important
observation for the main purpose of proving that the birth of a new
vision of Complexity accompanies the possibility of an experimental
detection of spontaneous collapses.

 Let us now shift
perspectives from the quantum manifestations of chaos and
 its
consequences to another paradox of quantum mechanics. As is well
known,
 the main hurdle for a satisfactory unification of quantum and
classical
 physics is the superposition principle, even if we assume
that the
 Correspondence Principle could provide the proper classical
limit of quantum
 phenomena. Let us assume that the time evolution of
two distinct, narrow
 wave packets A and B, each reproduces a
classical trajectory very well and
 the two trajectories are
macroscopically distinct. According to the linear
 nature of the
Shr\"{o}dinger equation if A is a solution of the equation, and B
 is
another solution of the equation, then so also is the linear
superposition of A and B. It is evident that the superposition of
two
 distinct outcomes, in this case two distinct classical
trajectories, is a
 concept foreign to classical mechanics and indeed
is essentially
 incompatible with our daily experience. The unfolding
of the dynamics of
 macroscopic bodies are known to be very
accurately described by Newton's
 equations. We actually discussed
this problem earlier in the form of the cat
 paradox. The resolution
of the paradox was offered by the theory of Zurek
 \cite{zurek} where
one takes into account the fact that there is no such
 thing in
nature as an isolated system, there is always a certain amount of
interaction with the environment. If we mimic the influence of the
external
 environment by white noise , a totally random process with
no memory, acting
 on the system of interest, then the correlations
between the two distinct
 classical trajectories (states of the cat's
well being) are lost. This
 resolution of the paradox makes it
impossible for an experimental
 observation, adopting the statistical
view of Gibbs, to assess the
 simultaneous existence of the two
trajectories. Thus, the environment,
 noise, has induced a collapse
of the wave function.

\subsection{Randomness is incompatible with traditional physics}

The resolution of the cat paradox by Zurek is very appealing, however,
it
 rests on the assumption that uncorrelated random processes can be
derived
 from within the ordinary laws of physics. This is a
perspective shared by
 the overwhelming majority of physicists, and
one of its earliest
 implementations was given by the celebrated
\emph{golden rule} of Fermi, see
 for example Zwanzig
\cite{zwanzig}. The golden rule is related to the
 possibility of
turning the coherent nature of a transition process in
 quantum
mechanics into an incoherent process strictly because of the large
number of particles in the system and the correspondingly large number
of
 quantum states participating in the transition. The uncritical
acceptance of
 this viewpoint has turned it into a scientific dogma
or prejudice rather
 than a description of what actually occurs.
 
This prejudice spread from the time of Fermi to the present in many
different guises, the Pauli master equation, the van Hove master
equation,
 see \cite{zwanzig,zwanzig2} for a complete discussion,
and
 transport processes based on the assumption of linear response,
see Bianucci
 et al. \cite{marco} for a modern treatment of these
ideas. Scientists have
 been quite satisfied because the statistical
predictions resulting from
 these theories, which are supposed to be
quantum, are completely consistent
 with experimental
observations. However, a careful analysis of all these
 theories by
Giovannetti et al. \cite{giovannetti} reveals a crack in the
foundation of physics in that they are all, in one way or another,
based on
 the Markov approximation. This dependence on the Markov
assumption is so
 fundamental to the physical theories on which our
understanding of the world
 is based that we are forced to discuss
this technical point of analysis at
 least at an intuitive level.

 Let us begin then by following a trajectory that is given a value
at the
 time {\it t}=0. Once the initial condition has been specified
the
 trajectory, determined by Newton's laws of motion, is fixed. The
trajectory
 is completly disconnected from the past, the future of
the system is only
 dependent on this initial state. This aspect of
determinism is obscured if
 rather than studying a trajectory, we
limit ourselves to considering a
 projection of this trajectory, much
like mistaking the shadows on the wall
 of Plato's cave for the
actual lives of people. If we are considering
 projected rather than
full trajectories, then two distinct projected
 trajectories can
depart from the same initial condition, at least in so far
 as can be
determined in the projected space. To realize that the two
 distinct
projected trajectories are a genuine manifestation of ordinary
classical mechanics, the observer has either to look at the full
trajectories or at the whole history of the trajectories. In examining
the
 full trajectories the observer will become convinced that in the
full phase
 space, the two trajectories never intersect, or if they
intersect once then
 they intersect infinitely often, that is, the
orbit is unstable (homoclinic
 orbit). In studying the whole history
of the trajectories the observer can
 remain at the projected level,
only to realize that the two trajectories
 departing from the ''same
initial condition'' actually are characterized by
 totally different
histories. Thus, we arrive at the surprising conclusion
 that the
deterministic character of the theory adopted, including quantum
mechanics if the process of measurement in excluded, is reflected in
their
 non-Markov character. By their non-Markov character we mean to
say
 that a projected representation must bear a significant
dependence on the
 past, and that the future time evolution of the
system does not depend only
 on the conditions of the system at the
moment of the observation, but it
 also depends on the history of the
system. This is not to be considered a
 special property of
statistical processes that make them different from the
deterministic nature of classical and quantum mechanics. On the
contrary,
 this significant dependence on the past is the mark of the
deterministic
 nature of the physical laws driving the time evolution
of the whole Universe.
 
 Ironically, all physicists claiming that
classical physics naturally emerge from quantum, adopt the Markov
approximation. Also Zurek claims that the current physical paradigms
are sufficient to account for all the fundamental problems concerning
the derivation of thermodynamics from classical mechanics, which in
turn is derived from quantum mechanics. Thus, a Markov process is a
statistical process whose time evolution is fixed only by the initial
condition, so that its evolution is totally independent of the
past. This property is shared by the deterministic evolution of the
entire universe, but when it becomes a statistical property of a
projection of the universe, it cannot be true. The Markov assumption
seems to be incompatible with a rigorously quantum treatment. Thus, if
it leads to plausible, and realistic results, it is a sign that
unknown physical laws drive the universe We think that these laws
introduce randomness into nature, thereby making it legitimate to
adopt the Markov condition.

 The Markov assumption produces an
exponential decay of correlations in the physical descriptions of
reality. However, Fonda et al. \cite{fonda} and Lee \cite{lee} proved
mathematical theorems establishing that exponential decay is
incompatible with both classical and quantum mechanics. Thus, the
Markov assumption seems to be a subterfuge, providing an illusion of
settling the fundamental problems of modern physics without any need
for additional hypotheses. Ironically, the repeated use of the Markov
approximation has been shown to be equivalent to a departure from
traditional physics.

 Thus, we see that the Markov
approximation is a consequence of a previous
 assumption, that being
the observation that the study of the entire universe
 is too complex
for us to address all at once. This observation implies that
 we
examine only a projected part of the universe. This decision itself
entails certain subjective elements and even if it lead to the
resolution of
 the quantum-related problems discussed above it would
still result in the
 Second Law of Thermodynamics being a consequence
of our limited knowledge of
 the universe, rather than being an
objective aspect of reality. The Markov
 approximation is therefore
inconsistent with the physical laws that the
 advocates of this
approach claim to be a complete representation of the
 universe.

 How is it possible that such a fundamentally incorrect assumption
can provide such a wealth of accurate predictions? A formal answer to
this question has been given by Giovannetti \emph{et al}
\cite{giovannetti} who define the conditions for a genuine source of
randomness to produce the Markov approximation with no significant
departure from the predictions made using traditional physics. On the
basis of the results of Giovannetti \emph{et al.} we wish to make the
following plausible conjecture: \emph{All the sources of complexity
examined so far are actually channels for the amplification of
naturally occurring randomness in the physical world.}

 This
randomness, must not be confused with algorithmic complexity. It is
a
 genuine property of nature independent of any experimental
observation. If
 the algorithmic complexity is so high as to result,
according to the
 arbitrary Markov approximation, in a very short
correlation time, then the
 spontaneous fluctuations might have the
effect of making the Markov
 approximation genuine. This is expected
to result in predictions slightly
 different from those of ordinary
quantum mechanics, but for practical
 purposes it might not have any
relevent consequences. If the ''subjective''
 source of randomness is
extremely strong, an even infinitesimally small
 genuine source of
stochasticity has the effect of making the system diffuse
incoherently. This is the reason why the scientists who assume
incoherent
 behavior without introducing objective randomness find
experimental
 vindication of their predictions. They obtain the right
answers but for the
 wrong reasons.

\subsection{Information approach to complexity}

We have pointed out that the concept of randomness as a consequence of
a
 lack of information is not totally satisfactory, and that there is
a need
 for a new concept of objective randomness, perhaps in the
form of a new
 principle of physics. This new physical principle
should be only a slight
 modification of traditional quantum
mechanics in which it is supplemented by
 the inclusion of a genuine
element of randomness \cite{GRW}. On the other
 hand, we cannot
easily establish the intrinsic nature of this randomness. Is
 it the
familiar Wiener process as Ghirardi, Rimini and Weber \cite{GRW}
claim? The Wiener precess is ordinarily assumed to be an idealization
of
 physical processes satisfactorily described by known physical
laws. We have
 seen, however, that this cannot be the complete story
since the white noise
 postulated by Zurek cannot be derived from
traditional quantum mechanics. In
 principle we cannot rule out the
possibility that the Wiener process
 introduced by \cite{GRW} to
correct ordinary quantum mechanics might have a
 deterministic origin
similar to that generated by chaos, although produced
 by some still
unkown deterministic mechanism.
 
 The existence of objective
randomness seems to be in conflict with the
 recent comments of
Landauer \cite{land}, who considers the universe itself
 to be a
computer with finite memory. This view of the world would imply that
the fluctuations produced by round-off errors in ordinary computers
would
 have a correspondence in nature resulting in fluctuations
being embedded in
 the fabric of reality. Thus, the Markov
approximation incompatible with
 either ordinary quantum or classical
mechanics, might be produced by the
 round-off errors of the
universe. This picture would also resolve the
 fundamental question
swirling around the foundation of the derivation of
 thermodynamics
from mechanics, and of classical physics from quantum as
 well. There
are several indications that round-off errors are
 indistinguishable
from genuine fluctuations, and that these fluctuations
 produce a
crossover from anomalous to ordinary statistical mechanics \cite
{renato,elena}, although at very large times, if the intensity of
these fluctuations is very weak.
 
 We see that if a still unknown
principle of statistics, requiring that
 nature is fundamentally
random and irreversible, then the unsatisfactory
 aspects of the
current definitions of complexity are resolved. This is true
 in
spite of the ambiguity in the meaning of randomness in this new
context.
 This is where the physics paradigm suitably extended may
play a crucial role
 in the development of measures of complex
phenomena.

\section{Conclusions: Objective Randomness inducing a New Vision of Complexity}

The discussion of the earlier sections is enough for us to reach a
conclusion fitting the Penrose's view about "why a new physics is
needed to understand the mind" \cite{penrose3}. However, in the last
few years there have been many new results, on which many of the
papers of these Proceedings are based, which are also suggesting a new
vision of complexity, which, hopefully, affords convincing answers to
many of the question discussed in this paper. In addition to those
mentioned in the earlier sections, other groups are also looking for a
picture of reality where randomness is already present at the
fundamental level.

Let us quote some relevant cases. An interesting proposal has been
made \cite{ord} for a realistic setting for Feynman paths. This is an
attempt at a realistic interpretation of the amplitudes, rather than
probabilities, in the Feynman interpretation of quantum mechanics with
the path-integral formalism. This new formulation rests on the
dynamics of a pair of entwined trajectories. The particles move on
entwined-pair trajectories in space time therefore generating the
impression of unitary time evolution, with dynamic rules, though, that
are as random as the random walker prescriptions of classical
mechanics. This is in a sense the reverse of the assumption implicitly
made by the advocates of decoherence, whose philosophy would lead us
to conclude that wave-function collapses, and with them the second
principle of thermodynamics itself, are an illusion of observers
forced by their human limitation to look at a limited portion of the
Universe. The authors of Ref. \cite{ord} conjecture that from their
theory a realistic interpretation of the wave-function collapse might
emerge. This is quite possible, due to the fact that the new physics
that they propose is essentially random and non-unitary.

Another approach to quantum mechanics moving from thermodynamics, with
the second law regarded as being a fundamental law of nature, rather
than an illusion of the human observer, has been proposed by El
Naschie\cite{mohammed1}. El Naschie proves that the Cantorian space
can serve as a geometrical model for a spaceÐtime support of the
thermodynamical approach. Additional work at uncovering some
unsuspected connections between the abstract algebra of wild topology
and high energy physics has been more recently found by the same
author\cite{mohammed2} Using the same perspective, the three Nicolis
\cite{nicolis} explained the two-slit delayed experiment without using
the Wheeler interpretation of the "observer participancy", setting
doubts on the independent existence of the Universe.

However, at the end of this paper devoted to looking for a
satisfactory definition of complexity, it is convenient to discuss the
consequence that a new physics might have on this specific issue. The
main problem with the work of Ref.\cite{giovannetti} is that the
conclusions might be more satisfactory from a philosophical point of
view, since the resulting diffusion equation, with the characteristics
of normal diffusion, is not the mere result of a contraction
procedure, equivalent to interpreting the second principle as a human
illusion, but the second principle is true, independently of the
existence of an observer. However, from a practical point of view the
advocates of de-coherence theory might conclude that the same result
is obtained with simpler calculation, and, consequently, applying the
Ockham principle\footnote{We are referring to William of Ockham, a
well known philosopher of the 14th century.  The medieval rule of
parsimony, or principle of economy, frequently used by Ockham came to
be known as Ockham's razor. The rule, which said that plurality should
not be assumed without necessity (or, in modern English, keep it as
simple as you can), was used to eliminate many pseudo-explanatory
entities.}, is true. For this reason, it is important to mention the
work of ref. \cite{luigi} that yields a remarkable result, this being
a different experimental result, according to the perspective
adopted. Another way to express the same conclusion is as follows. As
pointed out in earlier sections, the de-coherence wisdom rests on the
division of the Universe into two parts, the system of interest and
its environment. If the environment is the source of uncorrelated
fluctuations, the resulting Markov equation yields results that from a
statistical point of view are equivalent to those where real
collapses, and events, unpredictable events, take place. If we move
from this safe condition to a condition where the bath is responsible
for correlated fluctuations the statistical equivalence of the two
pictures is not longer guaranteed. At the time of this writing, the
research work on these delicate issues is not yet completed. However,
there are strong indications that the breakdown of the equivalence
between density and trajectories noticed by the authors of
Ref. \cite{mauro} is provoked by the occurrence of aging
\cite{angelo,gerardo}. This brings us back to the conception of
complexity as LSM.

The vision of LSM emerging from the dynamic model of
Refs.\cite{jacopo,massi,gerardo}, according to the authors of
Ref.\cite{thetwobuiatti's}, has a biological relevance and represents
a vision that, without conflicting with that of the Prigogine
school\cite{prigoginenouvellealliance}, affords additional arguments
to support the view that life is not foreign to nature, as
misrepresented by the conventional equilibrium statistical
mechanics. Even in the absence of a flow of energy from outside, we
can notice a natural tendency to the emergence of properties, such as
aging, that are conventionally attributed to living systems. We have
to notice that this vision of complexity emerges from the dynamic
approach of Ref.\cite{marco} extended to the case where long-range
correlation and memory are present. In this case the transition from
dynamics to thermodynamics is infinitely slow thereby suggesting that
this condition as a new state of matter, the earlier mentioned concept
of LSM.

This dynamic approach yielding the vision of LSM, on the other hand,
is at the basis of new techniques of analysis of time
series\cite{giulia} , which are currently used with success to assess
the complexity of the systems, from which these time series are
generated. As we have earlier mentioned the Kolmogorov complexity is
not computable, and these techniques, directly or indirectly related
to the concept of a Kolmogorov complexity, yield a computable measure
measure of complexity. It is interesting to remark that, although
these techniques are accurate, distinguishing with their help biotic
from a-biotic systems remains a challenging
issue\cite{stromatolite1,stromatolite2} . It is interesting to notice
that the field of complexity is reversing the current
perspective. While, as pointed out by Gunter\cite{gunter}, explaining
why rocks and life emerged at the same time, in the geological scale,
is a challenging issues for ordinary physics (this meaning for us,
essentially, ordinary statistical mechanics), from within the field of
complexity it is rather becoming challenging the distinction between
biotic and a-biotic systems, given the widespread tendency in nature
to establish long-range correlations.

Due to the importance of the vision of complexity suggested by the
paper of Ref. \cite{jacopo} before ending this paper is convenient to
devote some more comments to it. It is based on the concept of
L\'{e}vy walk, a process characterized by infinitely extended time
memory, and so time non-locality, which is slowly converted into a
L\'{e}vy flight, namely, as pointed out by the authors of
Ref.\cite{seshadri} into space non-locality. However, it takes an
infinite time for this transition to occur. Throughout this transition
process, lasting for an infinite time, the dynamic process is multi
scaling, rather than mono scaling. Thus, we are led to conclude that
the condition of scaling, a quite mono-scaling condition, departing
from the ordinary scaling of Brownian motion, is not an indication of
complexity. It is rather the inscription on a grave signaling that the
system was complex when it was alive. In fact, an exact mono-scaling
condition indicates that the Markov condition has been recovered, this
meaning that many uncorrelated and \emph{objective} jumps occurred,
Notice that this condition of thermal death, occurring without the
influence of environmental noise, which probably would make Gaussian
the resulting death, is an idealization of reality. However, this
idealization serves the desireable purpose of illustrating the dynamic
perspective of LSM. In this ideal condition, the system would age
forever without ever dying. Furthermore, no simple generalization of
diffusion equation is known, for a fair representation of this
process, thereby really implying the breakdown of the simplicity
condition.  What about a quantum derivation of LSM? We note that the
work of \cite{gerardo} refers to a real experiment, on the so called
blinking quantum dots (see the important paper of Jung, Barkai and
Silbey \cite{barkai} for details on this fundamental aspect). We are
inclined to believe that the jumps from the light to the darkness
state and vice-versa are triggered by the spontaneous GRW
collapses. To have a non-Poisson statistics for these collapses
we probably need to generalize the work of Tessieri \emph{et
al.}\cite{tessieri} to the non-Ohmic condition.  The authors of
Ref.\cite{tessieri} studied the case when the de-coherence of the
system of interest is produced by the interaction with a bath of
bosons, undergoing the GRW collapses. The calculation must be extended
to the case of a non-Ohmic bath: an interesting research program. 

In conclusion, we have to acknowledge that there are significant
attempts at reconciling general relativity to quantum
mechanics \cite{nottale,reconcile1,reconcile2,reconcile3,reconcile4,reconcile5}
using fractal geometry, namely, one of the theoretical ingredient of
complexity. Furthermore, as earlier remarked, the assumption of
randomness as an essential ingredient of the new
physics \cite{ord,mohammed1,mohammed2,nicolis} makes it natural to
perceive the second principle as real rather than as an illusion, as
it is subtlety implied by de-coherence theory.  We think that all
these authors are doing remarkable work to properly address the
challenge of Gunter \cite{gunter} who correctly perceives quantum
mechanics, general relativity and quantum mechanics, as three
different theories, with no connections. We stress that within this
context the dynamic approach to complexity, moving from the earlier
work of Ref. \cite{marco}, is producing some specific benefits,
although at more limited level of establishing a relation between
dynamics and thermodynamics, with two major results. The first is the
discovery of a promising direction to project experiments aiming at
turning a philosophical controversy about randomness and wave-function
collapses into a real scientific issue \cite{luigi}. The second is the
proposal \cite{gerardo} of a new view of complexity as a state of
transition from dynamics to thermodynamics, denoted as LSM, with the
important effect of abolishing the perspective of ordinary statistical
mechanics that would make life foreign to physics.

\emph{Acknowledgments} PG acknowledges support from ARO, through Grant
DAAD19-02-0037.

\end{document}